\documentclass[12pt]{article}
\usepackage[utf8]{inputenc}
\usepackage[margin=1in]{geometry}
\usepackage{setspace}
\usepackage[T1]{fontenc}
\usepackage{newtxtext,newtxmath}
\usepackage{microtype}

\usepackage{amsmath,amssymb}
\usepackage{graphicx}
\usepackage{booktabs}
\usepackage{siunitx}
\usepackage{enumitem}
\usepackage{xcolor}
\usepackage{bbding}
\usepackage{hyperref}
\usepackage{multirow}
\usepackage{array}
\usepackage{makecell}
\usepackage{pifont}
\usepackage{tabularx}
\usepackage{url}

\newtheorem{definition}{Definition}
\newcommand{\ignore}[1]{}
\hypersetup{colorlinks=true, linkcolor=black, urlcolor=blue, citecolor=blue}
\usepackage[
  backend=biber,
  style=numeric-comp,
  sorting=none,
  citetracker=true
]{biblatex}
\usepackage{etoolbox}
\makeatletter
\pretocmd{\footnotesize}{\normalsize}{}{}
\makeatother
\defbibheading{notes}{\section*{Notes}}

\title{Cognitive Warfare:\\Definition, Framework, and Case Study}

\author{
Bonnie Rushing\\
Laboratory for Cybersecurity Dynamics \\
Department of Computer Science \\
University of Colorado Colorado Springs\\ US Air Force
\and
William Hersch\\
Military and Strategic Studies Department\\
US Air Force Academy
\and
Shouhuai Xu\\
Laboratory for Cybersecurity Dynamics \\
Department of Computer Science \\
University of Colorado Colorado Springs
}

\begin{document}
\maketitle

\begin{abstract}
Cognitive warfare has emerged as a central feature of modern conflict, yet it remains inconsistently defined and difficult to evaluate. Existing approaches often treat cognitive operations as a subset of information operations, limiting the ability to assess cognitive attacker-defender interactions or determine when advantage has been achieved. This article proposes a unified definition of cognitive warfare, introduces an interaction framework grounded in the OODA loop, and identifies measurable attributes associated with cognitive superiority. To illustrate the use of the framework, a notional case study demonstrates how these concepts can be applied to assess cognitive attacks and defenses in a contested environment. Thus, the framework provides joint force leaders and analysts with a practical foundation for understanding, comparing, and evaluating cognitive warfare campaigns.
\end{abstract}

\noindent\textit{Disclaimer: “The views expressed are those of the author and do not reflect the official policy or position of the US Air Force, Department of War, or the US Government.”}

\section{Introduction}
\label{sec:intro}

Contemporary competition increasingly targets the cognitive dimension of conflict. Rather than seeking decisive effects through physical force alone, adversaries aim to shape perception, manipulate belief, degrade trust, and influence decision-making at scale \cite{JointPub3-04,DOD-IO-Strategy-2023}. These activities operate largely below the threshold of armed conflict, even more so than cyber operations, yet they can produce strategic effects by altering how individuals and organizations observe their environment, orient to information, decide on courses of action, and act. As a result, cognitive advantage has become a central component of military effectiveness and national power.

Despite growing recognition of this challenge, cognitive warfare remains inconsistently defined and poorly evaluated. This is evidenced by the 2026 NDAA, which mandates that the Department of War propose a systematic definition \cite[sec. ``Narrative Intelligence and Cognitive Warfare'']{SenateReport119-39}. Indeed, existing doctrine and policy often subsume cognitive effects within broader information operations or influence activities, emphasizing message dissemination, narrative competition, or engagement metrics \cite{Wardle2017InformationDisorder,Paul2020RAND,Rushing}. While these perspectives capture important elements of the problem, they provide limited insight into whether cognitive attacks actually degrade decision quality or confer sustained advantage. Moreover, many existing models implicitly treat cognitive influence as a unidirectional activity, focusing on attacker techniques while underemphasizing defender adaptation, resilience, and learning \cite{Hutchins2022CognitiveWargaming}.

This article argues that cognitive warfare is best understood as an interactive struggle over human cognition, in which attackers and defenders continuously adapt in pursuit of advantage. Drawing on the Observe–Orient–Decide–Act (OODA) loop of Col John Boyd \cite{Osinga2007ScienceStrategy}, cognitive warfare can be framed as a competition to disrupt, delay, or distort an opponent’s cognitive processes while protecting one’s own \cite{JointPub3-04,Kahneman2011ThinkingFastSlow}. Within this framework, success is not measured solely by visibility or reach, but by the extent to which cognitive effects translate into degraded judgment, altered behavior, or loss of trust over time.
Specifically, our investigation is centered on the following research questions (RQs):
\begin{description}
\item[\textbf{RQ1:}] What attributes are essential for modeling and characterizing cognitive warfare?
\item[\textbf{RQ2:}] How should cognitive warfare be rigorously defined in a manner that distinguishes it from related concepts such as information operations and influence activities?
\item[\textbf{RQ3:}] How should we define cognitive superiority?
\end{description}

\subsection{Contributions}

The article makes three contributions. First, the article introduces a novel framework for understanding and characterizing cognitive warfare, dubbed the Multi-Horizon Cognitive OODA framework. The framework has three features: (i) It models cognitive warfare as a sustained attacker–defender competition over decision advantage, rather than a unidirectional messaging activity. (ii) It utilizes the OODA loop as a starting point, but goes much beyond by introducing and incorporating the notion of temporal horizon to describe the short-term vs. long-term effects of cognitive attacks and defenses. (iii) It is also consistent with doctrinal framing of the cognitive dimension and with decision-centric treatments of influence and cognition. 

Second, the article demonstrates the usefulness of the framework via a notional case study of cognitive attacks and defenses in a contested information environment. The case study shows how the framework provides a shared conceptual foundation for supporting joint force leaders and analysts in understanding when cognitive advantage has been gained or lost, and how cognitive warfare can be more effectively planned, assessed, and countered. 

Third, the article discusses the implications of cognitive warfare for the Joint Force, such as: cognitive warfare should be institutionalized as a decision-centric operational concept; decision resilience is a key capability; simulation and wargames should incorporate cognitive stressors.

\subsection{Related Work}
\label{sec:related}

Prior work relevant to cognitive warfare spans (i) definitions of influence and cognition in conflict, (ii) models of attacker-defender interaction in the information environment, and (iii) efforts to evaluate the effectiveness of influence activities. 

\paragraph{Prior Studies on Definitions of Cognitive Warfare and Influence.}
Existing definitions of cognitive warfare and related concepts vary widely in scope and emphasis. US policy and joint doctrine sources describe the cognitive dimension as the domain of human perception, decision-making, and action, often situating it within the broader information environment \cite{JointPub3-04,DOD-IO-Strategy-2023}. NATO and allied analyses extend this view, framing cognitive warfare as a deliberate effort to exploit cognitive vulnerabilities in individuals and societies to achieve strategic advantage \cite{DuCluzel2020CognitiveWarfare,NATOstratComCOE2023Vol11}.
Academic literature offers a broader view, drawing on psychology, social sciences, and information disorder research to examine how beliefs, trust, and narratives can be manipulated \cite{Wardle2017InformationDisorder,Kahneman2011ThinkingFastSlow}. However, these definitions often diverge in whether they emphasize \textit{mechanisms} (e.g., persuasion, manipulation), \textit{effects} (e.g., belief change, behavioral response), or \textit{levels of analysis} (individual versus population). As a result, cognitive warfare is frequently discussed without a shared, effects-based definition that clearly distinguishes it from information operations or strategic communication. By contrast, we propose a systematic definition of cognitive warfare.

\paragraph{Prior Studies on Attacker–Defender Interaction Models.}
A second body of work examines attacker-defender dynamics in the information environment, including wargaming, threat modeling, and influence operations analysis, and the closely related cyber domain \cite{DBLP:conf/hotsos/Xu14a,XuBookChapterCD2019,DBLP:conf/ccs/Xu20}. Studies on information environment wargames and cognitive wargaming highlight the importance of adaptation, learning, and competition between adversaries \cite{Paul2020RAND,Hutchins2022CognitiveWargaming}. These efforts demonstrate the value of modeling interaction rather than treating influence as a static activity.
Nevertheless, many existing cognitive attack-defense models remain either attacker-centric or structurally static, focusing on tactics and messaging while giving limited attention to defender cognition, resilience, and feedback loops. Even when the OODA loop is referenced, it is often applied conceptually rather than operationally, leaving gaps in how cognitive effects propagate through decision-making processes over time \cite{Wilden2023Benchmarking}. By contrast, this article presents a conceptual foundation that can be leveraged to build mathematically sound cognitive attack-defense dynamics models, similar to cybersecurity dynamics models \cite{DBLP:conf/hotsos/Xu14a,XuBookChapterCD2019,DBLP:conf/ccs/Xu20}.

\paragraph{Prior Studies on Metrics and Evaluation of Cognitive Effects.}
Evaluating success in cognitive warfare presents a persistent challenge. Much of the existing evaluation literature relies on proxy measures such as engagement, reach, or exposure, particularly in studies of misinformation and influence campaigns \cite{Roozenbeek2020HarmonySquare,Basol2020GoodNewsBadNews}. While these metrics are useful for understanding dissemination and susceptibility, they do not directly indicate whether cognitive attacks degrade decision quality or confer lasting advantage.
Recent work on psychological sophistication, social engineering, and training effectiveness suggests promising directions for more meaningful measurement, including the assessment of cognitive vulnerability, adaptation, and learning \cite{XuIEEEAccessSychologicalQuantification2024,Salas2009TrainingScience}. However, these approaches are rarely integrated into a unified framework capable of supporting comparative evaluation or determining cognitive superiority in contested environments. By contrast, this article introduces a suite of novel metrics to measure the effectiveness of cognitive attacks and defenses.

\subsection{Paper Outline}
The rest of the article is organized as follows. Section \ref{sec:framework} presents the framework. Section \ref{sec:casestudy} presents our case study. Section \ref{sec:conclusion} concludes the article.

\section{Conceptual Framework for Cognitive Warfare}
\label{sec:framework}

This section introduces a conceptual framework for cognitive warfare grounded in Boyd’s OODA loop.
The framework has three novel features.
First, the framework models cognitive warfare as a sustained attacker–defender competition over decision advantage, rather than a unidirectional messaging activity. 
Second, we distinguish cognitive warfare effects by \emph{temporal horizon}: (i) \emph{acute} (short-horizon) effects that manifest within minutes to days and are observable as immediate degradation of OODA performance (e.g., misperception, delayed decisions, reduced confidence); and (ii) \emph{chronic} (long-horizon) effects that accumulate over weeks to years and gradually shift the target's priors, trust calibration, and identity frames, namely that chronic effects ``re-parameterize'' acute OODA loops by shaping what is noticed, what is believed to be credible, and which actions appear legitimate or feasible.
These horizons enable us to systematically analyze how cognitive effects emerge, accumulate, and translate into relative decision superiority under sustained contest. Thus, we call this the Multi-Horizon Cognitive OODA framework. 
Third, the framework is consistent with doctrinal framing of the cognitive dimension and decision-centric treatments of influence and cognition \cite{JointPub3-04,DOD-IO-Strategy-2023,Kahneman2011ThinkingFastSlow,Wardle2017InformationDisorder,RushingXu2025}. 
Specifically, joint doctrine characterizes the cognitive dimension in terms of how individuals and groups perceive, process, and act on information, and it emphasizes that operational success depends on shaping adversary decision-making while protecting one’s own \cite{JointPub3-04,DOD-IO-Strategy-2023}. 
Our framework operationalizes this doctrinal intent by (i) mapping contestable cognitive functions to OODA stages (observe, orient, decide, act), (ii) treating cognitive warfare as an adaptive attacker--defender interaction rather than a one-way messaging activity, and (iii) making evaluation explicitly decision-centric (e.g., decision latency, decision error, trust calibration) rather than visibility-centric \cite{JointPub3-04,Kahneman2011ThinkingFastSlow,Wardle2017InformationDisorder}.

\begin{figure}[!htbp]
    \centering
\includegraphics[width=1\linewidth]{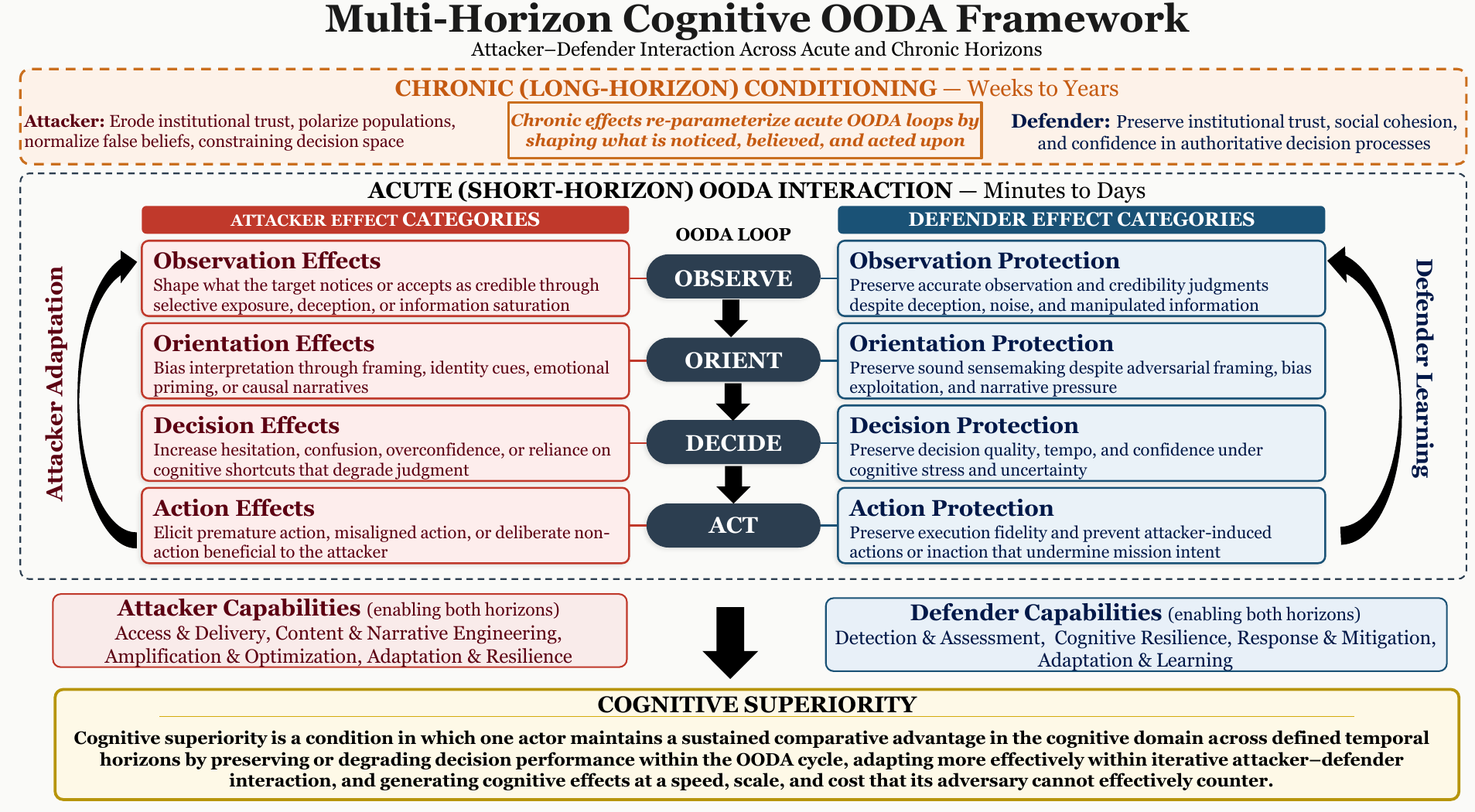}
\caption{The proposed framework for describing, understanding, and characterizing cognitive warfare. Each OODA loop stage represents a contested function in which the cognitive attacker (red) seeks to disrupt, delay, or distort the defender's cognitive processes, while the cognitive defender (blue) seeks to preserve decision advantage. Feedback loops reflect the adaptive nature of both sides. Cognitive superiority is achieved when one actor (i.e., attacker or defender) maintains sufficient advantage to perform OODA loop actions with reduced interference across short (acute) and long (chronic) horizons.}
\label{fig:ooda}
\end{figure}

Figure \ref{fig:ooda} highlights the framework. In what follows we elaborate its core tenets: cognitive warfare attributes definition (answering RQ1), cognitive attacks definition, cognitive defenses definition, the cognitive attack-defense effects mapping to the OODA loop, cognitive warfare definition (answering RQ2), and cognitive superiority definition (answering RQ3).

\subsection{Defining Attributes to Characterize Cognitive Warfare (Answering RQ1)}
\label{sec:attributes}

To enable systematic definition and evaluation, we characterize cognitive warfare using three families of attributes that apply symmetrically to both attackers and defenders: \emph{objectives} (what effects an actor seeks to produce or prevent), \emph{capabilities} (what an actor can do under contest to achieve or deny those effects), and \emph{cost and efficiency} (the resources required to sustain advantage). 

This three-family structure is consistent with (i) doctrinal emphasis on decision advantage and effects in the cognitive dimension \cite{JointPub3-04,DOD-IO-Strategy-2023}, (ii) empirical findings that exposure/engagement proxies are insufficient to infer cognitive impact \cite{Wardle2017InformationDisorder,Roozenbeek2020HarmonySquare,Basol2020GoodNewsBadNews}, and (iii) a long tradition of modeling adversarial contests by specifying objectives, capabilities, and resource asymmetries \cite{Paul2020RAND,DBLP:conf/hotsos/Xu14a,DBLP:conf/ccs/Xu20}.

\paragraph{1. Objectives (Effects over short and long horizons).}
Objectives specify the intended or avoided cognitive effects and incorporate two embedded attributes: \emph{decision degradation} and \emph{temporal horizon}. The \textbf{decision degradation} attribute reflects measurable changes in OODA performance, including increased decision latency, increased decision error, reduced confidence, or induced misaligned action within mission-relevant time windows \cite{JointPub3-04,Kahneman2011ThinkingFastSlow}. The \textbf{temporal horizon} attribute distinguishes between \emph{short-horizon (acute)} objectives, which manifest within minutes to days as immediate OODA disruption, and \emph{long-horizon (chronic)} objectives, which accumulate over weeks to years as persistent shifts in trust calibration, belief alignment, and interpretive priors that shape subsequent decision cycles \cite{DuCluzel2020CognitiveWarfare,NATOChiefScientist2026CognitiveWarfare,Wardle2017InformationDisorder}. In both cases, objectives should be stated with explicit target populations, time windows, and decision consequences to support comparison and assessment.

\paragraph{2. Capabilities (Achieving objectives under contest).}
Capabilities specify the mechanisms an actor can employ to achieve (or deny) cognitive objectives in an adaptive attacker--defender contest and incorporate two embedded attributes: \emph{interaction/adaptation} and \emph{scalability/amplification} \cite{Paul2020RAND,Hutchins2022CognitiveWargaming}. The \textbf{interaction/adaptation} attribute reflects the ability to adjust tactics, detect adversary activity, reduce recovery time, and learn under sustained contest. The \textbf{scalability and amplification} attribute reflects the ability to generate, constrain, or mitigate cognitive effects rapidly and at scale, including access and delivery, narrative engineering, and optimization mechanisms \cite{Wardle2017InformationDisorder,Cialdini2009Influence}. For attackers, these capabilities include access, narrative engineering, amplification, and adaptation. For defenders, they include detection and assessment, cognitive resilience, response and mitigation, and organizational learning \cite{Roozenbeek2020HarmonySquare,Salas2009TrainingScience}. Capability assessment emphasizes performance under pressure, whether an actor can adapt faster than an opponent, reduce recovery time following disruption, or limit amplification before acute effects translate into measurable decision degradation.

\paragraph{3. Cost and efficiency (Sustaining advantage).}
Cost and efficiency constitute a distinct attribute reflecting the \textbf{asymmetric cost structure} of cognitive warfare. Cognitive contests often allow relatively low-resource actions to generate disproportionate cognitive effects, particularly in digitally mediated environments \cite{JointPub3-0,NATOChiefScientist2026CognitiveWarfare}. Accordingly, evaluation requires assessing not only whether objectives are achieved, but whether they are achieved efficiently and sustainably under prolonged competition. These attributes capture the resource burden required to generate or deny cognitive effects at scale over time and enables comparative assessment of which actor can sustain cognitive advantage.

Together, these three attribute families provide a practical and measurable basis for characterizing cognitive warfare as a dynamic interaction rather than a unidirectional influence process \cite{Paul2020RAND,Hutchins2022CognitiveWargaming}, thereby addressing \textbf{RQ1}. Table \ref{tab:attribute-families} summarizes the attribute families in a symmetric attacker-defender structure, making explicit the framework’s decision-centric focus on OODA performance and its distinction between short-horizon disruption and long-horizon conditioning.

\begin{table}[htbp]
\centering
\caption{Symmetric attribute structure for evaluating cognitive warfare. The three attribute families organize decision-centric and temporal elements within an attacker--defender contest.}
\label{tab:attribute-families}
\footnotesize
\renewcommand{\arraystretch}{1.35}
\begin{tabularx}{\textwidth}{
>{\raggedright\arraybackslash\bfseries}p{2.1cm}
>{\raggedright\arraybackslash\bfseries}p{2.5cm}
>{\raggedright\arraybackslash}X
>{\raggedright\arraybackslash}X}
\toprule
\textbf{Attribute Family} & \textbf{Embedded Attribute} & \textbf{Attacker Expression} & \textbf{Defender Expression} \\
\midrule

Objectives
& Decision degradation (acute)
& Increase decision latency, error, hesitation, or induce misaligned action within mission-relevant time windows.
& Preserve OODA integrity, tempo, and execution fidelity under cognitive pressure. \\

& Temporal conditioning (chronic)
& Shift trust calibration, belief alignment, and interpretive priors over time.
& Preserve institutional trust, calibrated belief, and cohesion across extended horizons. \\

\addlinespace

Capabilities
& Adaptation and interaction
& Adjust tactics, narratives, and delivery mechanisms under defender pressure.
& Detect, mitigate, recover, and learn faster than adversary adaptations. \\

& Scalability and amplification
& Amplify narratives rapidly and broadly to maximize acute and cumulative effects.
& Limit amplification, reduce exposure, and constrain spread before effects degrade decisions. \\

\addlinespace

Cost and Efficiency
& Asymmetric cost structure
& Achieve cognitive effects at scale with minimal resource expenditure.
& Deny or mitigate adversary effects without unsustainable resource burden. \\

\bottomrule
\end{tabularx}
\end{table}

\subsection{Cognitive Attacks Definition}

Unlike information activities that focus primarily on message dissemination or visibility, cognitive attacks are defined by their intended \emph{effects} on human cognition. In particular, they seek to degrade how actors observe, orient, decide, and act by exploiting cognitive biases, trust relationships, and mental shortcuts \cite{JointPub3-04,Kahneman2011ThinkingFastSlow}. Technology and information platforms function as enablers of scale and persistence, but the human cognitive process remains the primary target. This prompts us to adopt a recent definition.

\begin{definition}[Cognitive Attacks \cite{RushingXu2025}]
\label{definition:Cognitive Attacks}
Cognitive attacks are operations that target human minds, aiming to manipulate perceptions and beliefs, which may be weaponized and enhanced through technology and deceptive information, typically to affect the decision-making and actions of individuals or broader populations to gain advantage.
\end{definition}

Definition \ref{definition:Cognitive Attacks} emphasizes decision degradation and behavioral impact as the distinguishing outcomes of cognitive attacks, rather than information exposure alone. As such, cognitive attacks are most meaningfully evaluated by whether they measurably alter judgment, delay or distort decision-making, or induce actions (or inaction) that advantage the attacker over the relevant time horizon (short/acute vs. long/chronic) and level of analysis (individual, organizational, population). 

Inspired by how adversarial contests are specified in cybersecurity (objectives, capabilities, and adaptation under defense), we propose specifying cognitive attacks in terms of their \emph{objectives} and \emph{capabilities}, consistent with the attribute families introduced in Section~\ref{sec:attributes} \cite{DBLP:conf/hotsos/Xu14a,DBLP:conf/ccs/Xu20,Paul2020RAND}.
In particular, attacker objectives correspond to the \emph{objectives} family (decision degradation over acute horizons and conditioning over chronic horizons), while attacker capabilities correspond to the \emph{capabilities} family (delivery, narrative engineering, amplification, and adaptation under contest) \cite{JointPub3-04,Wardle2017InformationDisorder,Paul2020RAND}.

\subsubsection{Attacker Objectives}
\label{sec:attacker-objectives}

In the context of cognitive warfare, the attacker’s objective is not merely to disseminate information, but to
generate cognitive effects that translate into decision advantage \cite{JointPub3-04,DOD-IO-Strategy-2023,NATOChiefScientist2026CognitiveWarfare}.
To avoid conflating levels of war with time-scale, we specify attacker objectives by \emph{temporal horizon}:
(i) \emph{short-horizon (acute)} objectives that manifest within minutes to days as immediate degradation of OODA performance, and
(ii) \emph{long-horizon (chronic)} objectives that accumulate over weeks to years by shifting priors, trust calibration, and identity frames.
That is, attacker objectives should be specified with explicit time horizons, target populations, and intended
decision consequences to support evaluation and comparison.

\paragraph{Short-horizon (acute) attacker objectives.}
Short-horizon objectives aim to misdirect attention, degrade critical thinking, or induce specific behavioral
responses among operators, staffs, or commanders within mission-relevant time windows.
More specifically, short-horizon objectives include (i) \emph{observation (perception) effects}, shaping what the target notices or accepts as credible;
(ii) \emph{orientation effects}, biasing interpretation through framing, identity cues, emotional priming, or causal narratives;
(iii) \emph{decision effects}, increasing hesitation, confusion, overconfidence, or reliance on cognitive shortcuts; and
(iv) \emph{action effects}, eliciting behaviors beneficial to the attacker, including non-action
\cite{Wardle2017InformationDisorder,Kahneman2011ThinkingFastSlow,JointPub3-04}.

\paragraph{Long-horizon (chronic) attacker objectives.}
Long-horizon objectives often involve persistent erosion of institutional trust, polarization of populations,
or normalization of false beliefs that constrain political and military decision space over time
\cite{DuCluzel2020CognitiveWarfare,NATOstratComCOE2023Vol11}.

\subsubsection{Attacker Capabilities}

Attacker capabilities are the resources and mechanisms used to produce cognitive effects and sustain them under contest. Attacker capabilities are assessed not only by their existence but by their effectiveness under defender pressure. For instance, at the \emph{short-horizon}, whether the attacker can maintain or escalate cognitive effects across the OODA loop despite detection, counter-messaging, training, and other defensive measures \cite{JointPub3-04,DOD-IO-Strategy-2023}.
We organize capabilities into four categories that can be applied to achieve both \emph{short-horizon (acute)} and \emph{long-horizon (chronic)} objectives:

\begin{itemize}
\item \emph{Access and delivery} is the ability to reach target audiences through platforms, interpersonal channels, or compromised information pathways, including spoofed or deceptive sources \cite{Wardle2017InformationDisorder}. 

\item \emph{Content and narrative engineering} is the ability to craft persuasive or manipulative messages tailored to cognitive vulnerabilities, social identity cues, and situational stressors \cite{Cialdini2009Influence,Kahneman2011ThinkingFastSlow}. 

\item \emph{Amplification and optimization} includes tactics that increase visibility, perceived consensus, and salience (e.g., coordinated inauthentic behavior, timing strategies, and microtargeting), thereby influencing what targets observe and treat as socially validated \cite{Wardle2017InformationDisorder,Paul2020RAND}.

\item \emph{Adaptation and resilience}: the capability to adjust tactics, narratives, and delivery methods in response to defender mitigation, public awareness, or platform enforcement \cite{Paul2020RAND,Hutchins2022CognitiveWargaming}. The DISARM framework provides a structured taxonomy of adversary tactics and techniques that can be used to parameterize this capability set for red-team modeling and comparative analysis \cite{DISARM}.
\end{itemize}

\subsection{Cognitive Defense Definition}

In the context of cognitive warfare, defenses are not limited to counter-messaging or platform enforcement, but encompass the deliberate actions and capacities that prevent, absorb, or recover from adversary efforts to degrade decision-making. This prompts us to define:

\begin{definition}[Cognitive Defenses]
\label{definition:Cognitive Defenses}
Cognitive defenses are operations and capacities that protect human minds from adversarial efforts to manipulate perceptions and beliefs, preserving or restoring sound decision-making and action in pursuit of sustained advantage.
\end{definition}

Similarly, we specify cognitive defenses in terms of defender objectives (the cognitive effects the defender seeks to preserve across horizons) and defender capabilities (the mechanisms used to achieve those objectives under contest).

\subsubsection{Defender Objectives}
\label{sec:defender-objectives}

In general, cognitive defenses do not require eliminating, or preventing all exposure to, all adversary narratives. Defensive success is achieved when adversary activity fails to produce \emph{meaningful} degradation of defender performance
over the relevant horizon(s).
Symmetric to attacker objectives, defender objectives are specified by \emph{temporal horizon}:
(i) \emph{short-horizon (acute)} objectives that preserve OODA integrity and tempo under immediate cognitive pressure, and
(ii) \emph{long-horizon (chronic)} objectives that preserve institutional trust, social cohesion, and trust calibration over time.

\paragraph{Short-horizon (acute) defensive objectives.}
Preserve decision advantage by protecting the integrity and tempo of the defender’s OODA loop in mission-relevant time windows:
maintaining situational awareness, preventing confusion, and enabling timely and accurate decisions by operators, staffs, and commanders.
More specifically, the defender objective corresponds directly to the OODA loop stages and describes what the defender aims to protect:
\emph{observation protection},  \emph{orientation protection},  \emph{decision protection}, and  \emph{action protection}.

\paragraph{Long-horizon (chronic) defensive objectives.}
Preserve institutional trust, social cohesion, and confidence in authoritative decision processes over extended time horizons,
thereby reducing the attacker’s ability to constrain political or military decision space.

\subsubsection{Defender Capabilities}
\label{sec:defender-capabilities}

In general, defender capabilities go much beyond counter-messaging and include detection, resilience, adaptation, and learning across the cognitive domain.
Specifically, defender capabilities are the mechanisms that can be used to achieve defensive objectives. Defender capabilities are evaluated by their ability to reduce the effectiveness, persistence, or scalability of cognitive attacks while preserving decision speed and accuracy under sustained adversarial pressure. 
In parallel to attacker capabilities, we consider four kinds of defender capabilities:
\begin{itemize}
\item \emph{Detection and assessment} capabilities enable defenders to identify cognitive attacks, attribute intent, and assess potential impact. This includes recognizing manipulative tactics, coordinated influence activity, and indicators of cognitive stress or distortion within target populations. 
Effective detection supports timely defensive action but is insufficient on its own.

\item \emph{Cognitive resilience} capabilities aim to reduce susceptibility to manipulation by strengthening critical thinking, trust calibration, and awareness of adversarial techniques. For instance, research on psychological inoculation, prebunking, and training effectiveness demonstrates that exposure to tactics and techniques can improve resistance to cognitive attacks over time \cite{Roozenbeek2020HarmonySquare,Kiili2024Tackling,Salas2009TrainingScience}.

\item \emph{Response and mitigation} capabilities include counter-framing, authoritative signaling, and procedural safeguards that reduce the influence of adversarial narratives without amplifying them. For instance, these measures operate primarily at the orientation and decision stages of the OODA loop, helping defenders recover from partial cognitive compromise.

\item \emph{Adaptation and learning} capabilities allow defenders to adjust defensive strategies as adversaries evolve their tactics. Wargaming and iterative evaluation play a critical role in this process by enabling defenders to rehearse cognitive contests, test assumptions, and identify vulnerabilities before they are exploited in real-world operations.
\end{itemize}

\subsection{The Cognitive Attack-Defense Effects Mapping to the OODA Loop}

Since attacker's acute and chronic \textit{objectives} are different,
attacker's acute and chronic \textit{effects} are also different. The same applies to defender's acute and chronic effects.  Still, the two objectives are related to each other because chronic (long-horizon) objectives shift the baselines and priors of acute  effects.
Table~\ref{tab:ooda-mapping} describes the acute effects mapped to the OODA loop.

\begin{table}[!htbp]
\centering
\caption{Mapping Acute (Short-Horizon) Attacker and Defender Effects to the OODA Loop.}
\label{tab:ooda-mapping}
\small
\renewcommand{\arraystretch}{1.5}
\begin{tabularx}{\textwidth}{>{\bfseries\raggedright\arraybackslash}p{1.8cm} >{\raggedright\arraybackslash}X >{\raggedright\arraybackslash}X}
\toprule
\textbf{OODA Stage} & \textbf{Attacker Effect Category} & \textbf{Defender Effect Category} \\
\midrule
Observe
& \textbf{Observation Effects.} Shape what the target notices or accepts as credible through selective exposure, deception, or information saturation.
& \textbf{Observation Protection.} Preserve accurate observation and credibility judgments despite deception, noise, and manipulated information. \\
Orient
& \textbf{Orientation Effects.} Bias interpretation through framing, identity cues, emotional priming, or causal narratives.
& \textbf{Orientation Protection.} Preserve sound sensemaking despite adversarial framing, bias exploitation, and narrative pressure. \\
Decide
& \textbf{Decision Effects.} Increase hesitation, confusion, overconfidence, or reliance on cognitive shortcuts that degrade judgment.
& \textbf{Decision Protection.} Preserve decision quality, tempo, and confidence under cognitive stress and uncertainty. \\
Act
& \textbf{Action Effects.} Elicit behaviors beneficial to the attacker, including premature action, misaligned action, or deliberate non-action.
& \textbf{Action Protection.} Preserve execution fidelity and prevent attacker-induced actions or inaction that undermine mission intent. \\
\bottomrule
\end{tabularx}
\end{table}

On the other hand, \emph{chronic} (long-horizon) effects operate by shifting the baseline conditions under which those cycles run. Specifically, chronic cognitive effects re-parameterize OODA by shaping (i) what is treated as salient or credible at the \emph{Observe} stage, (ii) the interpretive priors and identity frames applied during \emph{Orient}, (iii) the confidence thresholds and risk perceptions that govern \emph{Decide}, and (iv) the perceived legitimacy and feasibility of options at \emph{Act} \cite{JointPub3-04,DuCluzel2020CognitiveWarfare,NATOChiefScientist2026CognitiveWarfare,Wardle2017InformationDisorder}. In this way, long-horizon conditioning increases the likelihood that short-horizon disruptions translate into measurable decision degradation under contest.

\subsection{Cognitive Warfare Definition (Answering RQ2)}

Building on the definitions of cognitive attacks and cognitive defenses, we now define cognitive warfare. 
Importantly, the definition is informed by the three attribute families introduced in Section~\ref{sec:attributes}. 
First, the \emph{objectives} family motivates an effects-based definition centered on decision degradation across acute and chronic horizons (rather than exposure or messaging alone). 
Second, the \emph{capabilities} family motivates modeling cognitive warfare as an interactive and adaptive contest, because both attackers and defenders employ scalable delivery, narrative engineering, mitigation, and learning under ongoing opposition. 
Third, the \emph{cost and efficiency} family motivates the emphasis on \emph{sustained} competition, since asymmetric resource structures often determine which actor can maintain pressure or resilience over time \cite{JointPub3-04,DOD-IO-Strategy-2023,Paul2020RAND,Hutchins2022CognitiveWargaming,Wardle2017InformationDisorder}.

\begin{definition}[Cognitive Warfare]
\label{definition:cognitive-warfare}
Cognitive warfare is a sustained and adaptive contest over human decision-making in which adversaries seek relative advantage by shaping or disrupting perception, interpretation, judgment, and action over time.
\end{definition}

Definition \ref{definition:cognitive-warfare} states that cognitive warfare, like cyber warfare, is inherently interactive because cognitive attackers and defenders continuously observe, respond, and adapt to one another across iterative OODA cycles. This attribute distinguishes cognitive warfare from predominantly unidirectional messaging models. Moreover, unlike related activities such as information operations or psychological operations, cognitive warfare is distinguished not by the tools employed, but by its primary target, effects, and evaluative criteria. Furthermore, we compare cognitive warfare with information operations, psychological operations, and influence activities across six distinguishing dimensions, depicted in Table \ref{tab:cw-comparison} and elaborated below.

\begin{table}[htbp]
\centering
\caption{Comparison between cognitive warfare and related concepts. Cognitive warfare is distinguished from information operations, psychological operations, and influence activities across six dimensions. Unlike related concepts that emphasize informational coordination, persuasion, or audience engagement, cognitive warfare targets human cognition directly, defines success through decision degradation, and is evaluated through decision-centric outcomes.}
\label{tab:cw-comparison}
\footnotesize
\renewcommand{\arraystretch}{1.35}
\begin{tabularx}{\textwidth}{>{\raggedright\arraybackslash\bfseries}p{2.25cm} >{\raggedright\arraybackslash}X >{\raggedright\arraybackslash}X >{\raggedright\arraybackslash}X >{\raggedright\arraybackslash}X}
\toprule
\textbf{Dimension} & \textbf{Cognitive Warfare} & \textbf{Information Operations} & \textbf{Psychological Operations} & \textbf{Influence Activities} \\
\midrule

Primary Target
& Human cognition: perception, orientation, decision, and action
& Information environment, systems, and channels
& Attitudes and beliefs of target audiences
& Public opinion, legitimacy, and political will \\

Central Objective
& Decision degradation under contest: altered judgment, delayed/distorted decisions, constrained courses of action
& Informational coordination: synchronize information-related activities across domains
& Persuasion/messaging effects: attitudinal or behavioral change in target audiences
& Shape perceptions to coerce, deter, legitimize, or undermine \\

Interaction Model
& Dynamic, adaptive attacker--defender contest with continuous feedback across OODA loops
& Coordination-centric: integrate and synchronize capabilities; feedback often indirect
& Predominantly unidirectional: message--audience focus with limited adaptive modeling
& Predominantly unidirectional: actor--audience focus; adaptation varies by campaign \\

Time Horizon
& \textbf{Multi-horizon:} \emph{short-horizon (acute)} OODA disruption (minutes--days) and
\emph{long-horizon (chronic)} conditioning (weeks--years) with cumulative effects
& Typically aligned to operation/campaign timelines; may be short or sustained depending on mission design
& Campaign-based messaging; commonly short-to-mid duration, though repeatable across campaigns
& Varies widely: short-term bursts through sustained, long-running narrative efforts \\

Level of Analysis
& \textbf{Independent of time horizon:} individual, unit, organization, institution, or population (effects may propagate across levels)
& Often operational/strategic planning focus, but effects may target any level depending on integrated activities
& Primarily population or subgroup (target audience) level; can include elite decision-makers
& Primarily population or decision-elite level; can also target communities and institutions \\

Evaluation Criteria
& Decision-centric outcomes: decision quality, tempo, confidence, trust calibration, and behavior under contest
& Operational effectiveness: mission objectives achieved via coordinated information-related activities
& Reach/exposure plus attitudinal/behavioral measures (surveys, compliance indicators, messaging effects)
& Reach/engagement plus perception and legitimacy indicators; attribution and impact often uncertain \\

\bottomrule
\end{tabularx}
\end{table}

First, \emph{the human mind is the primary contested domain}. Cognitive warfare directly targets perception, orientation, decision-making, and action, rather than information systems or physical infrastructure \cite{JointPub3-04,DuCluzel2020CognitiveWarfare}. While cyber and information capabilities may be used as enablers, success is ultimately determined by effects on human cognition.

Second, \emph{effects on decision-making are the central objective}. Cognitive warfare seeks to degrade judgment, distort situational understanding, increase hesitation or error, and constrain available courses of action \cite{Kahneman2011ThinkingFastSlow,RushingXu2025}. This emphasis on decision degradation differentiates cognitive warfare from activities focused primarily on persuasion, messaging, or audience reach.

Third, \emph{cognitive warfare is inherently interactive and adaptive}. Attackers and defenders continuously observe, respond, and adjust to one another’s actions, producing a dynamic contest rather than a one-directional influence effort \cite{Paul2020RAND,Hutchins2022CognitiveWargaming}. As a result, both offense and defense must be analyzed together to understand cognitive advantage.

Fourth, \emph{cognitive effects operate across multiple levels of analysis}. Cognitive warfare may target individuals and units at the tactical level, organizations and institutions at the operational level, or populations and decision elites at the strategic level \cite{Wardle2017InformationDisorder,NATOstratComCOE2023Vol11}. Effects at one level may propagate to others over time, particularly when trust and legitimacy are eroded.

Fifth, \emph{cognitive warfare unfolds over extended time horizons}. Unlike discrete cyber or kinetic actions, cognitive warfare often relies on persistence, repetition, and cumulative effects that shape beliefs and behavior gradually \cite{DuCluzel2020CognitiveWarfare,NATOChiefScientist2026CognitiveWarfare}. This temporal dimension complicates detection and assessment but is central to strategic impact.

Sixth, \emph{cognitive warfare is evaluable through decision-centric outcomes}. While exposure and engagement metrics may indicate activity, cognitive warfare is most meaningfully assessed by whether cognitive effects translate into measurable changes in decision quality, decision speed, trust, or behavior under contest \cite{JointPub3-04,DOD-IO-Strategy-2023}.

The preceding six dimensions collectively distinguish cognitive warfare as a distinct form of conflict centered on cognition itself, rather than merely an extension of information operations, psychological operations, or cyber activity, addressing \textbf{RQ2}.

\subsection{Defining Cognitive Superiority (Answering RQ3)}

The concept of cognitive superiority draws on the established notion of air superiority, which describes a condition in which one actor possesses sufficient advantage to operate without prohibitive interference \cite{JointPub3-0}. Air superiority does not imply absolute control or elimination of adversary capabilities, but rather a relative and context-dependent freedom of action. Adapting this logic to the cognitive domain provides an analytical foundation for defining advantage under sustained cognitive contest.

The preceding attributes provide a structural basis for defining \textit{cognitive superiority}. Intuitively, superiority is not determined by any single indicator, but by comparative performance across attributes over a defined period of contest. An actor demonstrates cognitive superiority when its decision performance remains resilient while its adversary experiences sustained degradation, when it adapts more rapidly to evolving tactics, and when it generates or denies cognitive effects at a speed, scale, and cost that the opponent cannot effectively counter. 
That s, superiority is therefore not static; it emerges from comparative adaptation speed, detection capability, and resilience under sustained contest.

\begin{definition}[Cognitive Superiority]
\label{definition:cognitive-superiority}
Cognitive superiority is a condition in which one actor maintains a sustained comparative advantage in the cognitive domain as manifested by the cognitive capabilities to achieve its cognitive objectives (across defined temporal horizons by preserving or degrading decision performance within the OODA cycle) at will and higher speed, with much higher cost-effectiveness than its opponent.
\end{definition}

Definition \ref{definition:cognitive-superiority} states that superiority occurs when attacker objectives across \emph{short- and long-horizons} persist despite defender actions and measurably degrade defender OODA performance (e.g., increased decision latency, increased decision error, or degraded trust calibration in authoritative channels). 
It anchors superiority in measurable OODA performance across short- and long-horizon effects, incorporates adaptive attacker–defender interaction, and accounts for speed, scalability, and cost efficiency under sustained competition. Cognitive superiority does not require control of perception or belief; it exists when an actor sustains relative decision advantage under contest. Superiority is lost when adversarial influence introduces sufficient cognitive friction to delay decisions, distort judgment, or constrain viable courses of action, even absent overt persuasion \cite{DOD-IO-Strategy-2023}.

Definition \ref{definition:cognitive-superiority} does {\em not} state that cognitive superiority require control of perception or belief; it exists when an actor sustains relative decision advantage under contest. An actor may retain superiority despite exposure to adversarial narratives so long as its OODA processes remain resilient and effective. Conversely, superiority is lost when adversarial influence introduces sufficient cognitive friction to delay decisions, distort judgment, or constrain viable courses of action, even absent overt persuasion or belief change. Unlike traditional influence metrics that emphasize reach or attitudinal shifts, cognitive superiority is evaluated through decision-centric outcomes: whether actors can observe accurately, orient coherently, decide confidently, and act effectively under pressure \cite{DOD-IO-Strategy-2023}. This framing provides a practical foundation for assessing cognitive advantage in dynamic attacker–defender interaction and motivates the following case study.

\section{Case Study: Cognitive Warfare in a Notional Contested Information Environment}
\label{sec:casestudy}

To demonstrate the usefulness of the  framework, we present a notional case study of cognitive warfare in a contested information environment. The purpose is not to replicate a specific real-world operation, but to illustrate how an analyst could specify attacker and defender objectives/capabilities, map acute effects to OODA stages, account for chronic conditioning, and evaluate cognitive superiority using decision-centric indicators.

\noindent\textbf{Case Study Notation.} 
To simplify presentation, we use compact labels throughout the case study:
AAO$\#$ (Attacker Acute Objective), 
ACO$\#$ (Attacker Chronic Objective), 
AC$\#$ (Attacker Capability), 
ACE$\#$ (Attacker Cost and Efficiency), 
DAO$\#$ (Defender Acute Objective), 
DCO$\#$ (Defender Chronic Objective), 
DC$\#$ (Defender Capability), and 
DCE$\#$ (Defender Cost and Efficiency).

\subsection{Scenario Settings}
\label{sec:scenario}

\paragraph{Operating context.}
A joint task force (JTF) is deployed to support stability operations in the fictional partner nation of \textit{Norland}. Norland’s government is negotiating a logistics access agreement that would enable allied basing and humanitarian throughput. A near-peer adversary (\textit{Red}) seeks to prevent ratification by degrading trust in the JTF and the Norland government, while also creating acute decision friction during a short, high-stakes operational window. We present out case study online at \url{https://sites.google.com/view/cognitivewarfarecasestudy/}.

\paragraph{Key actors and decision nodes.}
The defender (\textit{Blue}) includes: (i) the JTF commander and staff, (ii) Norland’s Ministry of Transport and Interior, (iii) local emergency managers, and (iv) a Norland public audience whose cooperation affects freedom of maneuver.
The attacker (\textit{Red}) includes: (i) influence operators, (ii) cyber-enabled access teams, and (iii) coordinated inauthentic networks used to amplify narratives.

\paragraph{Time horizons and campaign phases.}
We frame the contest across two horizons:
\begin{itemize}[leftmargin=*]
\item \textbf{Acute phase (72 hours).} Red triggers a surge of deceptive and saturating content timed to a logistics incident (real but ambiguous) to create immediate OODA disruption within the JTF/Norland decision cycle.
\item \textbf{Chronic phase (Weeks--Months).} Red conducts persistent conditioning to erode trust in institutions and increase belief receptivity to crisis narratives.
\end{itemize}

\paragraph{Critical decisions (examples).}
During the acute 72-hour window, Blue must rapidly decide: (D1) whether to pause convoy movement, (D2) whether to impose curfews, (D3) whether to publicly attribute the incident, and (D4) whether to continue the basing ratification timeline. These decisions create measurable opportunities for decision latency, decision error, and misaligned action under cognitive pressure.

\paragraph{Notional data streams available for assessment.}
The assessment cell has access to: (i) time-stamped command-and-control (C2) logs (decision timestamps, meeting cadence), (ii) incident reports and after-action notes, (iii) social media monitoring summaries (volume/velocity indicators), and (iv) limited public sentiment polling (coarse trust indicators). This supports decision-centric evaluation without relying solely on exposure metrics.

\subsection{Cognitive Attacker Objectives, Capabilities, Cost and Efficiency}
\label{sec:attacker-casestudy}

\subsubsection{Attacker Objectives (Acute and Chronic)}
\label{sec:attacker-objectives-cs}

We specify attacker objectives by horizon and by intended decision consequences.

\paragraph{Acute (short-horizon) attacker objectives.}
During the 72-hour window, Red seeks immediate OODA disruption:
\begin{itemize}[leftmargin=*]
\item \textbf{AAO1 (Observe):} Create information saturation and selective deception such that Blue/Norland decision-makers cannot rapidly establish shared situational awareness.
\item \textbf{AAO2 (Orient):} Induce attribution ambiguity and emotionally salient framing (fear/anger) to bias sensemaking toward worst-case interpretations.
\item \textbf{AAO3 (Decide):} Increase decision latency (delays, re-litigating assumptions) and increase decision error risk (premature pauses, misattribution, overreaction).
\item \textbf{AAO4 (Act):} Elicit misaligned action or non-action (e.g., unnecessary restrictions that reduce freedom of maneuver; delayed convoy movement that undermines confidence).
\end{itemize}

\paragraph{Chronic (long-horizon) attacker objectives.}
Over weeks--months, Red aims to shift interpretive priors and trust calibration in Norland:
\begin{itemize}[leftmargin=*]
\item \textbf{ACO1: Institutional trust erosion.} Reduce trust in Norland institutions and in allied presence, increasing baseline skepticism toward official statements.
\item \textbf{ACO2: Identity-frame polarization.} Strengthen identity cues (``us vs.\ them'') that bias orientation and increase narrative receptivity during crises.
\item \textbf{ACO3: Credibility inversion.} Increase the perceived credibility of non-authoritative channels relative to official channels, thereby re-parameterizing observation and orientation under uncertainty.
\end{itemize}

\subsubsection{Attacker Capabilities (Applied Under Contest)}
\label{sec:attacker-capabilities-cs}

Red leverages four capability categories (Section~\ref{sec:attacker-objectives}): access/delivery, narrative engineering, amplification, and adaptation.

\paragraph{AC1: Access and Delivery.}
Red maintains persistent access to key information pathways:
\begin{itemize}[leftmargin=*]
\item \textbf{Channel access:} seeded local-language outlets, influencer proxies, and impersonation-ready accounts.
\item \textbf{Targeted delivery:} microtargeted content aimed at emergency managers and civil society nodes.
\end{itemize}
Operationally, this capability is implemented through pre-positioned personas and cutouts, localized content pipelines, and routine seeding that normalizes Red-aligned sources prior to the acute window \cite{Wardle2017InformationDisorder,Paul2020RAND}.

\paragraph{AC2: Content and Narrative Engineering.}
Red prepares pre-scripted narrative variants tied to likely incidents:
\begin{itemize}[leftmargin=*]
\item \textbf{Narrative motifs:} ``foreign troops caused the incident,'' ``government is hiding the truth,'' ``aid is being diverted.''
\item \textbf{Cognitive levers:} fear priming, blame assignment, false dilemmas, and credibility hijacking via look-alike sources.
\end{itemize}
Operationally, this capability is implemented via rapid A/B testing of frames, reuse of proven motifs, and tailoring to identity cues and situational stressors to increase plausibility and salience \cite{Cialdini2009Influence,Kahneman2011ThinkingFastSlow,Wardle2017InformationDisorder}.

\paragraph{AC3: Amplification and Optimization.}
Red uses coordinated networks to maximize early salience:
\begin{itemize}[leftmargin=*]
\item \textbf{Velocity:} high burst rate in the first hours to shape what audiences observe as ``what everyone is talking about.''
\item \textbf{Perceived consensus:} synchronized reposting to simulate agreement and inevitability.
\end{itemize}
Operationally, this capability is implemented using coordinated inauthentic behavior (e.g., bot-assisted reposting and synchronized accounts), timing strategies, and cross-platform reseeding to sustain visibility during the acute decision window \cite{Wardle2017InformationDisorder,Paul2020RAND}.

\paragraph{AC4: Adaptation and Resilience.}
Red adjusts narrative variants in response to Blue mitigation:
\begin{itemize}[leftmargin=*]
\item \textbf{Pivoting:} shifting from ``it happened'' to ``they are covering it up'' when debunks appear.
\item \textbf{Reseeding:} reintroducing claims through new accounts after takedowns.
\end{itemize}
Operationally, this capability is implemented via playbooks that map likely defender responses to prepared pivots, plus rapid regeneration of delivery accounts and content variants when platforms enforce removal \cite{Paul2020RAND,Hutchins2022CognitiveWargaming,DISARM}.

\subsubsection{Attacker Cost and Efficiency}
\label{sec:attacker-cost-cs}

Red’s resource investment is modest relative to the potential operational disruption:
\begin{itemize}[leftmargin=*]
\item \textbf{ACE1: Low marginal cost per narrative variant} due to reuse of templates and automation-assisted content generation.
\item \textbf{ACE2: High leverage} via algorithmic amplification and social retransmission: limited initial seeding can yield large downstream effects.
\item \textbf{ACE3: Asymmetric burden transfer:} Blue must verify, coordinate, message, and manage second-order effects, often at higher cost and slower tempo.
\end{itemize}

\subsection{Cognitive Defender Objectives, Capabilities, Cost and Efficiency}
\label{sec:defender-casestudy}

\subsubsection{Defender Objectives (Acute and Chronic)}
\label{sec:defender-objectives-cs}

Blue’s defensive objectives are symmetric to Red’s attack objectives.

\paragraph{Acute (short-horizon) defensive objectives.}
\begin{itemize}[leftmargin=*]
\item \textbf{DAO1 (Observe):} Maintain accurate observation and credibility judgments despite attacker employment of access, deception, and saturation capabilities (AC1, AC3 in this case study), ensuring that shared situational awareness converges within mission-relevant time windows.

\item \textbf{DAO2 (Orient):} Preserve coherent sensemaking despite attacker employment of narrative engineering and emotional priming capabilities (AC2 in this case study), preventing attribution ambiguity and framing bias from fragmenting staff interpretation.

\item \textbf{DAO3 (Decide):} Preserve decision tempo and decision quality and avoid overreaction or underreaction despite attacker amplification and ambiguity-inducing tactics (AC2--AC3), including attempts to trigger precautionary pauses, premature attribution, or escalatory signaling.

\item \textbf{DAO4 (Act):} Preserve execution fidelity and prevent attacker-induced misaligned action or paralyzing non-action despite coordinated narrative surges and perceived consensus effects (AC3), ensuring that operational activities remain aligned with mission intent.
\end{itemize}

\paragraph{Chronic (long-horizon) defensive objectives.}
\begin{itemize}[leftmargin=*]
\item \textbf{DCO1: Preserve calibrated trust.} Sustain confidence in authoritative processes without requiring unanimous belief despite attacker sustained employment of credibility inversion and trust-erosion narratives (ACO1 and AC2 in this case study).

\item \textbf{DCO2: Maintain cohesion.} Reduce polarization and preserve shared interpretive priors needed for crisis response despite attacker identity-frame amplification and ``us versus them'' conditioning (ACO2 and AC3 in this case study).

\item \textbf{DCO3: Strengthen credibility routing.} Ensure decision-makers and the public know where and how to verify critical claims despite attacker sustained reseeding, impersonation, and salience manipulation tactics (AC1--AC4), thereby reducing long-term susceptibility during future acute events.
\end{itemize}

\subsubsection{Defender Capabilities (Applied Under Contest)}
\label{sec:defender-capabilities-cs}

\paragraph{DC1: Detection and Assessment.}
Blue maintains a small assessment cell responsible for identifying cognitive attack indicators, including narrative velocity spikes, coordinated posting patterns, anomalous source behavior, and sudden framing convergence across platforms. The cell integrates open-source monitoring, intelligence reporting, and incident logs to assess potential impact on OODA stages (e.g., observation distortion or orientation divergence). However, attribution confidence is limited, reporting pipelines are fragmented, and senior decision-makers receive mixed-quality or delayed assessments. As a result, detection does not consistently translate into timely orientation protection \cite{Wardle2017InformationDisorder,Paul2020RAND}.

\paragraph{DC2: Cognitive Resilience.}
Blue has implemented awareness training focused on misinformation recognition, credibility routing, and verification discipline. These programs draw on inoculation-style approaches and decision-science insights intended to reduce susceptibility to manipulative framing and cognitive shortcuts \cite{Roozenbeek2020HarmonySquare,Kiili2024Tackling,Salas2009TrainingScience}. However, training penetration is uneven across staff sections and partner-nation counterparts. During crisis conditions, time pressure and stress reduce adherence to verification protocols, increasing reliance on heuristics and prior beliefs. Thus, resilience exists but is inconsistently applied under acute pressure.

\paragraph{DC3: Response and Mitigation.}
Blue possesses the capability to issue authoritative statements, synchronize messaging with Norland officials, and deploy counter-framing to stabilize orientation. Response mechanisms include pre-drafted holding statements, designated spokesperson channels, and procedural safeguards for fact validation. Nevertheless, clearance processes, legal review, and interagency coordination introduce latency during the acute window. Additionally, counter-framing is implemented cautiously to avoid amplifying adversarial narratives, which can slow corrective salience \cite{Wardle2017InformationDisorder}. Consequently, mitigation reduces long-term narrative persistence but may not prevent short-horizon decision friction.

\paragraph{DC4: Adaptation and Learning.}
Blue conducts periodic tabletop exercises and cognitive wargaming sessions designed to stress-test decision processes under information contest \cite{Hutchins2022CognitiveWargaming}. After-action reviews identify vulnerabilities in credibility routing, coordination tempo, and decision thresholds. However, institutional learning cycles are slower than Red’s narrative pivoting tempo. While lessons are eventually integrated into procedures and training updates, adaptation is episodic rather than continuous. As a result, Blue improves over longer horizons but struggles to match attacker adaptation speed during acute phases.
\subsubsection{Defender Cost and Efficiency}
\label{sec:defender-cost-cs}

\paragraph{Defender Cost and Efficiency.}
Blue faces a structurally asymmetric cost profile in which defensive actions require coordination, verification, and sustained staffing.

\begin{itemize}[leftmargin=*]
\item \textbf{DCE1: Verification burden.} Establishing ground truth requires cross-organization coordination, forensic review, and legal clearance. This burden can be measured intuitively by time-to-confirmation, number of coordination touchpoints required, and personnel-hours expended per incident.

\item \textbf{DCE2: Coordination overhead.} Aligning public messaging and operational responses across JTF, Norland ministries, and partner entities increases latency. This can be measured by time from detection to synchronized public statement, number of approval layers, and divergence between internal and external communication timelines.

\item \textbf{DCE3: Sustained staffing costs.} Persistent monitoring, analytic tooling, and rehearsal cycles require dedicated personnel and infrastructure. This can be measured through sustained manpower allocation, training cycle frequency, analytic tool maintenance costs, and opportunity costs relative to other mission priorities.
\end{itemize}

\subsection{Evaluation of Cognitive Superiority}
\label{sec:eval-superiority}

We now evaluate the contest using the framework’s three attribute families (objectives, capabilities, and cost and efficiency) and a decision-centric interpretation of cognitive superiority (Definition~\ref{definition:cognitive-superiority}). The goal is to show how an analyst can reach a defensible superiority judgment \emph{without} collapsing evaluation into engagement-only metrics.

\subsubsection{OODA-Stage Acute Effects and Observable Indicators}

Table~\ref{tab:cs-indicators} provides a notional mapping from acute attacker effects to decision-centric observables that a JTF assessment cell could realistically track.

\begin{table}[htbp]
\centering
\caption{Notional acute (short-horizon) indicators for evaluating OODA disruption in the case study.}
\label{tab:cs-indicators}
\footnotesize
\renewcommand{\arraystretch}{1.25}
\begin{tabularx}{\textwidth}{>{\bfseries\raggedright\arraybackslash}p{1.6cm} >{\raggedright\arraybackslash}X >{\raggedright\arraybackslash}X}
\toprule
\textbf{Stage} & \textbf{Attacker Effect (Acute)} & \textbf{Decision-Centric Observable (Examples)} \\
\midrule
Observe
& Saturation + selective deception reduces shared situational awareness.
& Increased time-to-validate critical facts; conflicting ``ground truth'' artifacts; higher rework in common operating picture. \\
Orient
& Framing and emotional priming bias interpretation; attribution ambiguity spreads.
& Increased divergence in staff interpretations; repeated re-litigating of assumptions; elevated reliance on non-authoritative sources. \\
Decide
& Induced hesitation/overreaction increases decision latency and error risk.
& Longer time from incident notification to decision; higher number of decision reversals; deviation from pre-briefed decision thresholds. \\
Act
& Misaligned action or non-action undermines mission intent.
& Pauses that reduce freedom of maneuver; delayed convoys; inconsistent public guidance; operational objectives missed in time window. \\
\bottomrule
\end{tabularx}
\end{table}

\subsubsection{Chronic Conditioning and Its Interaction with Acute Disruption}
In this notional scenario, Red’s chronic objectives (AC1-AC3) shape the priors that feed acute OODA cycles. For example, if credibility inversion and trust erosion have progressed, then during the acute 72-hour incident, Blue’s authoritative corrections are slower to restore orientation coherence. This produces a compounding effect: chronic conditioning increases the probability that acute deception yields measurable decision degradation (e.g., longer latency and greater risk of misaligned action).

\subsubsection{Why the attacker achieves cognitive superiority in the notional scenario}
We argue Red achieves cognitive superiority in this case study because it outperforms Blue across the three attribute families:

\paragraph{(1) Objectives achieved (decision degradation over horizons).}
During the acute window, the defender experiences measurable degradation consistent with AAO1--AAO4: delayed convergence on shared situational awareness, contested interpretation, slowed decision tempo, and at least one misaligned operational pause. Over the chronic horizon, trust erosion and credibility inversion increase receptivity to Red narratives, raising the baseline difficulty of rapid restoration.

\paragraph{(2) Capabilities dominate under contest (adaptation + amplification).}
Red’s amplification velocity in the first hours shapes what is salient before Blue’s coordination and clearance processes can respond. Red also pivots narratives faster than Blue can synchronize counter-framing, sustaining contest pressure across observe--orient--decide stages. Blue’s detection exists but does not translate into sufficiently rapid mitigation to preserve decision tempo.

\paragraph{(3) Cost and efficiency favor the attacker (asymmetric burden transfer).}
Red’s marginal cost to generate narrative variants and reseed after takedowns is low relative to the operational and coordination costs imposed on Blue (verification, alignment, public communication, and second-order consequence management). Under sustained contest, this asymmetry allows Red to maintain pressure longer than Blue can sustain high-tempo mitigation without tradeoffs.

\subsubsection{Compact Comparative Summary}
Table~\ref{tab:cs-summary} summarizes the superiority judgment using your three-family structure. We draw two conclusions. First, given the stated objectives, capability performance under contest, and asymmetric cost structure, the attacker satisfies the core condition of cognitive superiority (Definition~\ref{definition:cognitive-superiority}): Red maintains sustained comparative advantage by achieving its cognitive objectives at higher speed and with higher cost-effectiveness than Blue, producing decision-centric degradation (latency, coherence loss, and action misalignment) during the mission-relevant window.

\begin{table}[htbp]
\centering
\caption{Notional superiority assessment summary using the three attribute families.}
\label{tab:cs-summary}
\footnotesize
\renewcommand{\arraystretch}{1.25}
\begin{tabularx}{\textwidth}{>{\bfseries\raggedright\arraybackslash}p{2.2cm} >{\raggedright\arraybackslash}X >{\raggedright\arraybackslash}X}
\toprule
\textbf{Family} & \textbf{Attacker (Red)} & \textbf{Defender (Blue)} \\
\midrule
Objectives
& Achieves acute OODA disruption and leverages chronic conditioning to increase receptivity and reduce recovery.
& Seeks to preserve OODA integrity, but experiences measurable latency and partial misalignment under pressure. \\
Capabilities
& High amplification velocity + rapid pivoting sustains pressure across OODA stages.
& Detection exists but response/coordination delays reduce mitigation effectiveness within the acute window. \\
Cost / Efficiency
& Low-cost narrative generation imposes high verification and coordination costs on Blue.
& Higher cost structure to verify, align, communicate, and manage second-order effects. \\
\bottomrule
\end{tabularx}
\end{table}

Second, the case study also clarifies what would be required for Blue to regain advantage: reduce decision latency via rehearsed verification procedures, pre-authorized messaging lanes, and faster cross-entity synchronization; increase resilience through targeted inoculation and credibility routing; and constrain amplification earlier to prevent acute salience dominance. These changes directly improve Blue capability performance under contest without requiring perfect suppression of adversary narratives.

\subsection{Discussion}
\label{sec:discussion}

\subsubsection{Further Discussion on the Case Study}

The case study demonstrates how the framework enables systematic analysis of cognitive warfare beyond message-centric or exposure-based models. Several observations emerge from the scenario.

First, the case study reinforces that cognitive warfare is fundamentally \emph{decision-centric}. The attacker’s acute objectives were not merely to shape narratives, but to introduce measurable friction into the defender’s OODA cycle. Decision latency increased, interpretive coherence decreased, and at least one operational pause reflected action-stage misalignment. These outcomes illustrate that cognitive effects become operationally relevant when they translate into degraded perception, delayed orientation, distorted judgment, or misaligned action.

Second, the scenario highlights the interaction between chronic conditioning and acute disruption. Chronic erosion of institutional trust and credibility inversion reshaped the interpretive priors that fed the defender’s acute decision cycles. As a result, authoritative corrective messaging required more time to restore orientation coherence. The case therefore illustrates how long-horizon conditioning re-parameterizes short-horizon OODA performance, increasing the probability that acute deception will produce measurable operational degradation.

Third, the case reveals the centrality of adaptation speed in determining relative advantage. Although the defender possessed detection and mitigation mechanisms, coordination latency and clearance procedures reduced mitigation effectiveness during the acute window. In contrast, the attacker demonstrated rapid narrative pivoting and amplification optimization. Cognitive superiority, therefore, emerged not from control of belief, but from comparative performance under sustained contest.

Fourth, the asymmetric cost structure of cognitive warfare significantly influenced contest dynamics. The attacker’s marginal cost to generate and reseed narrative variants was relatively low, while the defender incurred substantial verification, synchronization, and communication burdens. This cost asymmetry prolonged cognitive pressure and reduced defender agility. The case thus underscores that sustainable cognitive advantage depends not only on achieving effects, but on achieving them efficiently under prolonged competition.

Several limitations should be acknowledged. This case study is notional and does not draw on real operational data. Observable indicators such as decision latency and orientation divergence are presented illustratively rather than empirically measured. Additionally, the scenario simplifies complex interagency, allied, and sociopolitical dynamics that would influence real-world outcomes. Future research should empirically test the proposed attribute families against historical cases and simulation-based experiments to validate measurable thresholds for cognitive superiority.

Despite these limitations, the case study demonstrates the analytic utility of the Multi-Horizon Cognitive OODA framework and clarifies how cognitive warfare can be evaluated systematically rather than descriptively.

\subsubsection{Implications for the Joint Force}

The framework, as manifested through the case study, has the following implications for doctrine, force design, training, and operational assessment, which suggest that cognitive advantage is not achieved through information dominance alone but through resilient decision processes and adaptive organizational performance under sustained contest.

\paragraph{1. Institutionalize Cognitive Warfare as a Decision-Centric Operational Concept.}
Cognitive warfare should not be treated as a subordinate element of missions such as public affairs or cyber operations. Rather, it represents a sustained contest over decision advantage that directly affects mission outcomes. Doctrine should explicitly incorporate cognitive objectives, capability development, and superiority assessment within operational planning processes.

\paragraph{2. Prioritize Decision Resilience Training.}
Because cognitive superiority hinges on preserving OODA integrity under pressure, training programs should emphasize decision-making under information saturation, ambiguity, and adversarial framing. Cognitive resilience, including recognition of manipulative tactics, credibility routing, and stress-aware decision protocols, should be integrated into professional military education and live exercises.

\paragraph{3. Incorporate Cognitive Stressors into Simulation and Wargaming.}
Operational exercises should include realistic adversarial amplification dynamics and narrative pivoting behaviors to test adaptation speed and recovery time. Embedding cognitive stress within simulations enables commanders and staffs to identify vulnerabilities in orientation coherence and decision tempo before real-world exploitation occurs.

\paragraph{4. Reduce Organizational Latency.}
The case study illustrates how coordination overhead can degrade defensive performance. Cross-functional cognitive defense teams that integrate intelligence, operations, behavioral expertise, and communication functions may reduce response delays and improve mitigation effectiveness during acute windows.

\paragraph{5. Adopt Decision-Centric Measures of Effectiveness.}
Assessment frameworks should move beyond reach and engagement metrics to incorporate decision latency, decision error frequency, recovery time, and trust calibration indicators. Embedding these measures into after-action reviews and readiness evaluations will enable systematic assessment of cognitive superiority.

\section{Conclusion}
\label{sec:conclusion}

This article introduced the Multi-Horizon Cognitive OODA framework, including a novel definition of cognitive warfare and cognitive superiority, to model cognitive attacker–defender interaction as a dynamic contest over decision advantage. By distinguishing acute (short-horizon) disruption from chronic (long-horizon) conditioning and organizing analysis around objectives, capabilities, and cost and efficiency, the framework provides a systematic basis for evaluating cognitive superiority.

A notional case study demonstrated the usefulness of the framework and how cognitive warfare manifests operationally when adversarial activities translate into measurable degradation of observation accuracy, orientation coherence, decision tempo, and action fidelity. It further illustrated that cognitive superiority does not require control of perception or belief. Rather, superiority emerges when one actor sustains comparative advantage in preserving or degrading OODA performance at greater speed and with greater cost-efficiency than its opponent.
As a result, decision resilience under contested information conditions will become a defining determinant of operational and strategic success.

Future work should empirically validate measurable indicators of cognitive superiority, develop quantitative models of cognitive attack–defense dynamics analogous to cybersecurity dynamics models, and integrate cognitive performance metrics into joint operational assessment frameworks.

\printbibliography[heading=notes]

@incollection{XuBookChapterCD2019,
author={Shouhuai Xu},
title={Cybersecurity Dynamics: A Foundation for the Science of Cybersecurity},
booktitle={Proactive and Dynamic Network Defense},
publisher={Springer},
volume={74},
year={2019},
pages="1--31",
}

@inproceedings{DBLP:conf/ccs/Xu20,
  author       = {Shouhuai Xu},
  title        = {The Cybersecurity Dynamics Way of Thinking and Landscape},
  booktitle    = {Proceedings of the 7th {ACM} Workshop on Moving Target Defense, MTD@CCS
                  2020, USA, November 9, 2020},
  pages        = {69--80},
  publisher    = {{ACM}},
  year         = {2020},
  doi          = {10.1145/3411496.3421225},
}

@inproceedings{DBLP:conf/hotsos/Xu14a,
  author       = {Shouhuai Xu},
  title        = {Cybersecurity dynamics},
  booktitle    = {Proceedings of the 2014 Symposium and Bootcamp on the Science of Security, HotSoS 2014, Raleigh, NC, USA, April 08 - 09, 2014},
  pages        = {14},
  publisher    = {{ACM}},
  year         = {2014},
  doi          = {10.1145/2600176.2600190},
}

@article{Rushing,
	author = {Bonnie Rushing},
	title = {Analysis of Media Influence on Military Decision-Making},
	year = {2024},
	journal = {The 19th International Conference on Cyber Warfare and Security},
volume={19}, number={1}, pages={308-316}
}

@misc{DISARM,
  author = {{DISARM Foundation}},
  title  = {DISARM Framework Explorer},
  year   = {2024},
  url    = {https://disarm.foundation/explorer},
  note   = {Accessed Aug.\ 31, 2025}
}

@article{RushingXu2025,
  author    = {Bonnie Rushing and Shouhuai Xu},
  title     = {Characterising Cyber Cognitive Attacks},
  journal   = {RUSI Journal},
  volume={171}, 
  number={2},
  month = {3},
  year      = {2026}
}

@ARTICLE{XuIEEEAccessSychologicalQuantification2024,
  author={Longtchi, Theodore Tangie and Rodriguez, Rosana Montañez and Gwartney, Kora and Ear, Ekzhin and Azari, David P. and Kelley, Christopher P. and Xu, Shouhuai},
  journal={IEEE Access}, 
  title={Quantifying Psychological Sophistication of Malicious Emails}, 
  year={2024},
  volume={},
  number={},
  pages={1-1},
  doi={10.1109/ACCESS.2024.3514603}
}

@article{Kiili2024Tackling,
  author  = {Kiili, Kristian and Siuko, Jenni and Ninaus, Manuel},
  title   = {Tackling misinformation with games: a systematic literature review},
  journal = {Interactive Learning Environments},
  year    = {2024},
  pages   = {1--16},
  doi     = {10.1080/10494820.2023.2299999},
  url     = {https://www.tandfonline.com/doi/full/10.1080/10494820.2023.2299999}
}

@report{Paul2020RAND,
  author  = {Paul, Christopher and Wong, Leonard and Bartels, Elizabeth},
  title   = {Opportunities for Including the Information Environment in U.S. Marine Corps Wargames},
  institution = {RAND Corporation},
  year    = {2020},
  doi     = {10.7249/RR4406},
  url     = {https://www.rand.org/pubs/research_reports/RR4406.html}
}

@inproceedings{Wilden2023Benchmarking,
  author  = {Wilden, David and Nasim, Mehrdad and Williams, Paul},
  title   = {On Benchmarking and Validation in Wargames: A Scoping Review and Agenda for Future Research},
  booktitle = {European Conference on Cyber Warfare and Security (ECCWS)},
  year    = {2023},
  pages   = {1--10},
  url     = {https://www.researchgate.net/publication/371703801}
}

@article{Basol2020GoodNewsBadNews,
  author    = {Basol, Melisa and Roozenbeek, Jon and van der Linden, Sander},
  title     = {Good news about bad news: Gamified inoculation boosts confidence and cognitive immunity against fake news},
  journal   = {Journal of Cognition},
  volume    = {3},
  number    = {1},
  pages     = {1--9},
  year      = {2020},
  doi       = {10.5334/joc.91}
}

@article{Roozenbeek2020HarmonySquare,
  author    = {Roozenbeek, Jon and van der Linden, Sander},
  title     = {Breaking Harmony Square: A game that “inoculates” against political misinformation},
  journal   = {Harvard Kennedy School Misinformation Review},
  volume    = {1},
  number    = {8},
  year      = {2020},
  doi       = {10.37016/mr-2020-47}
}

@book{Kahneman2011ThinkingFastSlow,
  author    = {Kahneman, Daniel},
  title     = {Thinking, Fast and Slow},
  publisher = {Farrar, Straus and Giroux},
  year      = {2011}
}

@article{NATOstratComCOE2023Vol11,
  author    = {{NATO Strategic Communications Centre of Excellence}},
  title     = {Defence Strategic Communications, Volume 11, Autumn 2022},
  journal   = {Defence Strategic Communications},
  volume    = {11},
  year      = {2023},
  doi       = {10.30966/2018.RIGA.11.1},
  note      = {Riga: NATO StratCom COE}
}

@article{Salas2009TrainingScience,
  author  = {Salas, Eduardo and Tannenbaum, Scott I. and Kraiger, Kurt and Smith-Jentsch, Kimberly A.},
  title   = {The Science of Training and Development in Organizations: What Matters in Practice},
  journal = {Psychological Science in the Public Interest},
  year    = {2012},
  volume  = {13},
  number  = {2},
  pages   = {74--101}
}

@misc{DOD-IO-Strategy-2023,
  author       = {{U.S. Department of Defense}},
  title        = {{Department of Defense Information Operations Strategy}},
  year         = {2023},
  month        = {5},
  howpublished = {\url{https://media.defense.gov/2023/May/31/2003231168/-1/-1/0/DoD-Information-Operations-Strategy-2023.pdf}},
  note         = {Emphasizes the cognitive dimension as central to future competition and warfare.}
}

@book{JointPub3-04,
  author       = {{Joint Chiefs of Staff}},
  title        = {{Joint Publication 3-04: Information in Joint Operations}},
  year         = {2022},
  month        = {1},
  publisher    = {{Joint Chiefs of Staff}},
  address      = {Washington, D.C.},
  note         = {Defines the cognitive dimension of the information environment as the domain of human perception, decision, and action.}
}

@book{Cialdini2009Influence,
  author    = {Cialdini, Robert B.},
  title     = {Influence: Science and Practice},
  edition   = {5},
  publisher = {Pearson Education},
  year      = {2009},
  address   = {Boston, MA}
}

@report{Wardle2017InformationDisorder,
  author      = {Wardle, Claire and Derakhshan, Hossein},
  title       = {Information Disorder: Toward an Interdisciplinary Framework for Research and Policy Making},
  institution = {Council of Europe},
  year        = {2017},
  address     = {Strasbourg, France},
  url         = {https://rm.coe.int/information-disorder-toward-an-interdisciplinary-framework-for-re/168076277c}
}

@techreport{Hutchins2022CognitiveWargaming,
  author       = {Hutchins, Brett and Peck, Michael},
  title        = {Cognitive Wargaming: Understanding and Countering Information Influence},
  institution  = {NATO Strategic Communications Centre of Excellence},
  year         = {2022},
  note         = {Riga, Latvia}
}

@report{NATOChiefScientist2026CognitiveWarfare,
  author       = {{NATO Science \& Technology Organization}},
  title        = {Cognitive Warfare: NATO Chief Scientist Research Report},
  institution  = {NATO Science \& Technology Organization},
  year         = {2026},
  url          = {https://www.nato.int/content/dam/nato/webready/documents/sto/chief-scientist-report-cognitive-warfare.pdf},
  note         = {Accessed 31 Jan 2026}
}

@report{DuCluzel2020CognitiveWarfare,
  author       = {du Cluzel, François},
  title        = {Cognitive Warfare},
  institution  = {NATO Allied Command Transformation (Innovation Hub)},
  year         = {2020},
  url          = {https://innovationhub-act.org/wp-content/uploads/2023/12/20210122_CW-Final.pdf},
  note         = {Accessed 31 Jan 2026}
}

@report{SenateReport119-39,
  author       = {{U.S. Senate Committee on Armed Services}},
  title        = {{National Defense Authorization Act for Fiscal Year 2026}},
  institution  = {U.S. Government Publishing Office},
  number       = {S. Rep. No. 119-39},
  year         = {2025},
  url          = {https://www.congress.gov/committee-report/119th-congress/senate-report/39/1},
  note         = {119th Cong., 1st sess. See section ``Narrative Intelligence and Cognitive Warfare.''}
}

@book{Osinga2007ScienceStrategy,
  author    = {Osinga, Frans P. B.},
  title     = {Science, Strategy and War: The Strategic Theory of John Boyd},
  publisher = {Routledge},
  year      = {2007},
  address   = {London}
}

@book{JointPub3-0,
  author    = {{Joint Chiefs of Staff}},
  title     = {{Joint Publication 3-0: Joint Operations}},
  year      = {2022},
  publisher = {{Joint Chiefs of Staff}},
  address   = {Washington, D.C.}
}

\subsection{Author biographies}
\textbf{Bonnie Rushing (SMSgt, USAF)} is the first enlisted member assigned to a doctoral program. She is currently pursuing a Ph.D. in Cybersecurity in the Laboratory for Cybersecurity Dynamics at the University of Colorado Colorado Springs through the Air Force Institute of Technology. Since enlisting in 2009, she has served as a Special Operations airborne linguist, signals intelligence team lead, and US Air Force Academy senior faculty member. She has flown 700+ flight hours aboard four C-130 variants, deployed across Latin America, and taught strategy, innovation, and joint operations wargaming. Her research focuses on social engineering, media association, and cognitive security, with multiple peer-reviewed publications and international-level awards. Her honors include the Defense Meritorious Service Medal, BEYA and WOC Technology Leader Awards, and “40 Under 40” in STEM. More at \url{www.thebonnierushing.com}.

\textbf{Dr. William “Ox” Hersch, Ph.D. (Col, USAF, retired)} is a proven educator and practitioner of military and strategic studies with an eclectic background as an instructor, educator, leader, and operator. Originally from Pennsylvania, Dr. Hersch received his commission at Officer Training School, Maxwell Air Force Base, Alabama and served over 25 years on Active Duty. He is a veteran of both OPERATION ENDURING FREEDOM and OPERATION IRAQI FREEDOM, with over 1,800 total flight hours and 400 combat hours in the B-1B Lancer. Dr. Hersch is a graduated Political-Military Affairs Specialist, holds a Doctoral Degree in Military Strategy, and is practiced at developing, researching, teaching, and executing adaptive strategies across the spectrum of conflict.

\textbf{Dr. Shouhuai Xu, Ph.D.} is the Gallogly Chair Professor in Cybersecurity and the Founding Director of the Laboratory for Cybersecurity Dynamics, Department of Computer Science, University of Colorado Colorado Springs (UCCS). His research interest includes unifying cyber warfare, cognitive warfare, and electronic warfare into a single framework. His research has won several awards, including the 1st place at the 2019 Worldwide Adversarial Malware Classification Challenge organized by the MIT Lincoln Lab, 2023 USCYBERCOM Cyber REcon Analyst Award, 2024 USCYBERCOM Cyber REcon Hunter Award, 1st place at the 2024 US DoD NSIN/UC2 Cyber Innovators Challenge (Topic 3), and 2025 ACM SIGSOFT Distinguished Paper Award. His research has been funded by AFOSR, AFRL, ARL, ARO, DOE, NSA, NSF and ONR. He is the Technology Pillar Lead of an NSF Regional Innovation Hub (Phase 1), which focuses on making space infrastructures and systems resilient. He co-initiated the International Conference on Science of Cyber Security (SciSec) and serves as its Steering Committee Chair. He has served as Program Committee co-chair for several international conferences. He is/was an Associate Editor of IEEE Transactions on Dependable and Secure Computing (IEEE TDSC), IEEE Transactions on Information Forensics and Security (IEEE T-IFS), IEEE Transactions on Network Science and Engineering (IEEE TNSE), and Scientific Reports. He is a Distinguished Member of ACM (class 2024). More at \url{https://xu-lab.org}.

\end{document}